\documentclass[prb,aapm,mph,amsmath,amssymb,reprint,floatfix]{revtex4-2}

\usepackage{placeins}
\usepackage{latexsym}
\usepackage{graphicx}

\usepackage{subfigure}    

\usepackage{amsmath}
\usepackage{color}
\usepackage{xcolor}
\usepackage{soul}
\DeclareMathOperator{\sign}{sign}

\usepackage{relsize}
\usepackage{placeins}
\usepackage{amssymb}
\usepackage{amsthm}
\usepackage{amsfonts}
\usepackage{braket}
\usepackage{latexsym}
\usepackage{graphicx}
\usepackage{grffile}
\usepackage{hyperref}
\usepackage{multirow}
\usepackage{wrapfig}

\begin{document}
	
	%\date{}
	\title{}% Force line breaks with \\
	
	\author{John McFarland$^{(1)}$} 
	\email{swqecs@gmail.com}
	\author{Efstratios Manousakis$^{(1,2)}$}
	\email{manousakis@gmail.com}
	\affiliation{$^{(1)}$Department  of  Physics and National High Magnetic Field Laboratory, Florida  State  University,  Tallahassee,  FL  32306-4350,  USA}
	\affiliation{$^{(2)}$Department   of    Physics,  National and Kapodistrian University    of   Athens,
		Panepistimioupolis, Zografos, 157 84 Athens, Greece}
	%\ead{aryalniraj7@gmail.com} %,manousakis@magnet.fsu.edu}
	%\maketitle
	
	\title{Gradient Descent Optimization of Fermion Nodes in Diffusion Monte Carlo}
	%by gradient descent}

	\begin{abstract}

		We present a method for optimizing the location of the fermion ground-state nodes using a combination of diffusion Monte Carlo (DMC) and projected gradient descent (PGD).
		A PGD iteration shifts the parameters of an arbitrary node-fixing trial function in the opposite direction of the DMC energy gradient, while maintaining the cusp condition for atomic electrons. The energy gradient is calculated from DMC walker distributions
		by one of three methods we derive from an exact analytical expression. We combine our energy gradient calculation methods with different gradient descent algorithms and a projection operator that maintains the cusp condition. We apply this stochastic PGD method to trial functions of Be, Li$_2$, and Ne, all consisting of a single Slater determinant with randomized parameters, and find that the nodes dramatically improve to the same DMC energy as nodes optimized by variational Monte Carlo. Our method, therefore, departs from the standard procedure of optimizing the nodes with a non-DMC scheme such as variational Monte Carlo, Density function theory, or configuration interaction based calculation, which do not directly minimize the DMC energy.
		
	\end{abstract}
	
	\maketitle
	
	\section{Introduction}
	
	Diffusion Monte Carlo (DMC)\cite{doi:10.1063/1.431514,PhysRevLett.45.566,Ceperley1986}, also known as Green's function Monte Carlo or
	Projector Monte Carlo,  
	is a technique for projecting the many-body wavefunction to the ground state.
	It has been used to accurately tackle bosonic or
	non-frustrated quantum-spin systems (see Ref.~\onlinecite{RevModPhys.63.1} for
	list of References). In the case of fermionic systems, while in general one has
	to restrict oneself to the fixed-node approximation,
	where it has been accurately applied to nuclear physics\cite{PhysRevC.36.2026,10.3389/fphy.2020.00117,PhysRevC.70.054325}
	when the required initial trial function was accurate enough.
	Node relaxation has led to accurate results for
	the electron gas in the continuum\cite{PhysRevLett.45.566} and
	for electrons on a lattice\cite{PhysRevB.46.560,PhysRevB.47.11897,PhysRevLett.78.4609}. There are variants of the method, such as the
	constraint path\cite{PhysRevLett.74.3652,PhysRevLett.90.136401}      which have been successfully applied to condensed matter physics\cite{PhysRevX.10.031016,PhysRevLett.90.136401} problems.
	
	The DMC method samples the ground-state wavefunction $\Psi$ of $N$ particles with walkers. These are points in a $d N$-dimensional space ($d$ being the
	dimensionality of space) with a probability density equal to $\Psi \Psi_g$, where $\Psi_g$ is a guiding function used for importance sampling. DMC projects $\Psi$ to the ground state by propagating it along the imaginary time axis, which is done by changing the position and weight of each walker in a way that samples the Green's function (i.e. the matrix elements
	of the evolution operator). To prevent large variations in weights, walkers are regularly deleted, duplicated, or combined in a way that the initial and final $\Psi \Psi_g$ are proportional.
	
	For a fermionic $\Psi$, it is generally necessary to impose zero boundary conditions at an a priori chosen nodal surface. That is, the nodal surface of $\Psi$ is taken  to be the
	nodal surface of a trial function $\Psi_t$, which is  an approximate ground-state wavefunction produced by a non-DMC method. Typically $\Psi_t$ is a product of a symmetric Jastrow function\cite{Jastrow} factor, which describes correlations, and an anti-symmetric factor, which is a single or a combination of Slater determinants\cite{Slater} that describe  the nodes. It is standard to set $\Psi_t = \Psi_g$, in which case nodal boundary conditions are naturally imposed by imaginary-time propagation alone. In the present paper we distinguish $\Psi_t$ from $\Psi_g$ because we often use a node-less $\Psi_g$, in which case nodal boundary conditions are imposed by deleting those walkers that attempt to cross the nodes.
	
	DMC error is improved with a $\Psi_t$ that better approximates the exact ground state wavefunction. The only source of error that cannot be eliminated
	with the DMC propagation is the fixed node error, which is contained in just the anti-symmetric part of $\Psi_t$. The other sources of error, namely the finite time-step error, statistical error, and population control error, can be respectively controlled with a smaller time-step, a greater sample number, and a larger walker population. 
	
	This makes the choice of the nodes of $\Psi_t$ the fundamental approximation of DMC. The resulting fixed-node error typically ranges from about 82 to 435 meV per atom\cite{fixedNodeErrorRange}, and is generally controlled by increasing the complexity of $\Psi_t$ and better optimizing its parameters. Because DMC energy is an upper bound to the true ground state energy\cite{VariationalProperty}, a parameter optimization method that minimizes DMC energy is the most accurate. However, optimization methods in use will minimize some other quantity, such as the VMC energy, the Kohn-Sham energy, or the local VMC energy variance.\cite{EVarianceOptimization,PsiOptimization}
	
	The first attempt to optimize the parameters of $\Psi_t$ using DMC walker distributions alone was by Reboredo et al.\cite{Self-healing-DMC} They iteratively generated a new $\Psi_t$ by projecting the coefficients of its determinants (or pfaffians)  from the walker distribution of the previous $\Psi_t$. Mindful that it becomes expensive to evaluate a $\Psi_t$ with an increasing
	number of determinants, they also proposed a method to reduce the complexity of $\Psi_t$ by minimizing a cost function between the initial and reduced $\Psi_t$ that heavily penalized changes of the nodes.
	
	We propose to directly use the DMC energy as a ``cost function'' when optimizing the parameters of $\Psi_t$. This differs from using the VMC energy as a cost function.\cite{SteepestDescent} Our method iteratively performs PGD on $\Psi_t$ by shifting its parameters in a direction roughly opposite to the energy gradient, while maintaining any constraints (the electron-nuclear cusp condition in our examples) by projecting the parameters back to the surface satisfying the constraint with each iteration. Gradient descent has an advantage for large parameter number $N_p$ in that it does not require an estimate of $N_p \times N_p$ matrices, which is required by many others, e.g. the linear method.

	We calculate the gradient of each parameter iteration from walker samples using one of three methods that we label A, B, and C. The walker distributions are produced by DMC imaginary time propagation with nodes fixed by the parameters. We experiment with several gradient descent algorithms commonly used for machine learning. In order to keep our method self-reliant, during PGD we do not rely on a pre-optimized Jastrow function, which does not affect the fixed-node error, although it does affect the other errors.

	Our paper is organized as follows: In Sec. \ref{section:Method} we present the three methods for evaluating the derivative of the energy from the DMC walker distribution.
	In Sec. \ref{section:Implementation} we describe our implementation of gradient descent, including the DMC scheme used, practical issues, and the form of $\Psi_t$ and $\Psi_g$.  In Sec. \ref{section:Results} we present tests on the accuracy and speed of the three method used, results of PGD on trial functions of Be, Li$_2$, Ne, and F$_2$. In Sec. \ref{section:Discussion}, we discuss the advantages of our method and ways to improve it. We discuss parameter fluctuations in Appendix~\ref{section:Fluctuations}, and we describe the gradient descent algorithms used in Appendix~\ref{section:Algorithms}  and we compare them in Appendix~\ref{section:EffectiveAlpha}. 
	
	\section{Method}
	\label{section:Method}

	A PGD iteration shifts the nodes in a direction expected to lower the DMC energy $E$. The nodes are determined by the parameters $\theta_i$ of the anti-symmetric part of $\Psi_t$, and the shift of the nodes is determined by the gradient $\tfrac{\partial E}{\partial\theta_i}$. The gradient of a nodal surface is calculated with walker samples produced by DMC imaginary-time propagation of $\Psi \Psi_g$, with $\Psi$ the fixed-node ground state and $\Psi_g$ a guiding function. Thus PGD iterations change the nodal surface, while DMC iterations generate data for the gradient of a given the nodal surface.

	Provided the $\theta_i$'s are not already at a local minimum, $E$ will be lowered by the following change of parameters
	\begin{equation}
	\label{eqn:GradientDescent}
	\theta_i \rightarrow \theta_i - a_i \tfrac{\partial E}{\partial\theta_i},
	\end{equation}
	provided that $a_i$ is positive and sufficiently small. However, the presence of stochastic error in our gradient means that only the expectation of $E$  will be lowered with sufficiently small $a_i$. If the $\theta_i$ are constrained  (e.g. from the cusp condition or symmetry) we modify Eq. \ref{eqn:GradientDescent} to 
	\begin{equation}
	\label{eqn:ProjectedGradientDescent}
	\theta_i \rightarrow P(\theta_i - a_i \tfrac{\partial E}{\partial\theta_i}),
	\end{equation}
	where $P$ is the projection operator that moves the parameters to the nearest point on the manifold that satisfies the constraint. Some gradient descent algorithms, e.g. ADAM, simulate momentum with friction by replacing $\tfrac{\partial E}{\partial\theta_i}$ with an exponential trailing average of current and past iterations.
	
	\subsection{Derivative of the energy}
	\label{section:EnergyDerivative}

	Central to our method is calculating the DMC energy gradient in parameter space. 	Working in atomic units, we derive an expression for ${\partial E \over \partial \theta_i} $, also derived by Berman.\cite{BoundaryEffects} To simplify the derivation, let us again set the ground state $\Psi$ to zero except for one nodal pocket, as justified in Sec. \ref{section:EnergyDerivative}. Now let us examine how the energy, given as
	\begin{equation}  
	\label{eqn:walkerEnergy}
	E = \frac{\bra{\Psi}{H} \ket{\Psi}}{\braket{\Psi|\Psi}},
	\end{equation}  
	changes due to an infinitesimal change $\delta\theta_i$ in  one of the
	parameters, which shifts the node and, thus,
	changes $\Psi  $  to   $\Psi  +
	\delta\Psi$.  By doing that the  change    in   energy $\delta{E}$ is given by
	\begin{eqnarray} 
	\delta     E     &=&     \frac{     \bra{\delta\Psi}{(H-E)}
		\ket{\Psi}}{\braket{\Psi|\Psi}}           +           \frac{
		\bra{\Psi}{(H-E)} \ket{\delta\Psi}}{\braket{\Psi|\Psi}} \nonumber
	\\
	&+&              \frac{             \bra{\delta\Psi}{(H-E)}
		\ket{\delta\Psi}}{\braket{\Psi|\Psi}} .
	\label{eqn:deltaE1} 
	\end{eqnarray} 	
	We convert the above expression to an integral over the 3N
	dimensional position vector ${\bf R}$ and integrate by parts to
	obtain
	\begin{equation}  
	\label{eqn:deltaE2}
	\delta  E   =  \frac{\int  d{\bf R}   \;  \;
		2\delta\Psi(H-E)\Psi            +            \delta\Psi
		(H-E)\delta\Psi}{\braket{\Psi|\Psi}}   .
	\end{equation}        
	The  $\delta\Psi(H-E)\Psi$ term  of  Eq.~\ref{eqn:deltaE2} is
	zero except at the original nodes, where the action of the Laplacian on   a  discontinuity   of  $\nabla\Psi$   produces  a   delta function. The term $\delta\Psi
	(H-E)\delta\Psi$   of  the above equation
	is second order in $\delta\theta_i$  except at the original and shifted
	nodes, where the action of the Laplacian on the discontinuities of
	$\nabla\delta\Psi$ produce  zeroth order delta functions. Thus,  we only
	need to consider the operation  of the Laplacian on these discontinuities at the original and shifted nodes
	to evaluate Eq.~\ref{eqn:deltaE2} up to first order in $\delta\theta_i$.
	
	\begin{wrapfigure}{l}{0.13\textwidth}
	\begin{center}
		\includegraphics[width=0.13\textwidth]{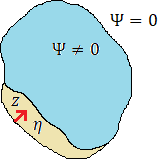}
		\end{center}
		\caption{Nodal-pocket. }
				\label{fig:Fig1}
	\end{wrapfigure}
	
	Let us now choose a  coordinate system with  the shape  of the
	nodes. Let $z$ be the coordinate perpendicular to the node, zero
	at the  initial node, and  pointing towards the direction where
	$\Psi\ne0$ (see Fig.~\ref{fig:Fig1} for an illustration). Let
	${\bf A}$  be  the  remaining $3N-1$  dimensional  coordinates  that
	parametrize the  nodal surface at $z=0$. Then $\Psi$ to first order is
	\begin{equation}
	\label{eqn:PsiAtNode}
	\Psi({\bf A},z) = g({\bf A})\textsf{R}(z),
	\end{equation}
	where $g ({\bf A} ) \equiv |\nabla\Psi({\bf A}, z=0^+)|$,
	and $\textsf{R}$ is the
	ramp function:
	\[  \textsf{R}(z) =
	\begin{cases}
	& z \quad \text{if } 0 \leq z \\ & 0 \quad \text{if } z < 0 \\
	\end{cases} .
	\] 
	To express $\delta\Psi$ near  the node, let us define $\eta({\bf A})$
	as  the displacement  of the  node in  the $z$  direction that
	results from  $\delta\theta_i$. Then,  to first order in $\Psi$,  $\delta\Psi$ is  the  difference between  the
	displaced and original wave function, that is
	\begin{equation}
	  \delta\Psi({\bf  A},z) =
          g({\bf A})\textsf{R}(z-\eta({\bf A})) - g({\bf A})\textsf{R}(z).
	\label{eqn:bondaryConditions}
	\end{equation}  
	The delta functions resulting from the Laplacian are then given by
	\begin{eqnarray}
	\label{eqn:bondaryConditions3}
	\nabla^2 \Psi({\bf A},z) = g({\bf A})\delta(z), \hskip 1.6 in
        \end{eqnarray}
        \begin{eqnarray}
	\nabla^2 \delta\Psi({\bf A},z)
	= g({\bf A})\delta(z-\eta) - g({\bf A})\delta(z). \hskip 0.6 in
	\end{eqnarray}
	Since only the delta functions contribute to first order, we leave only their contribution to  Eq.~\ref{eqn:deltaE2}, yielding
	\begin{equation}
	\delta  E   = -\frac{\int  d{\bf R}   \;  \;
		\delta\Psi({\bf A}, z)g({\bf A}) [\delta(z)
		+\delta(z-\eta)]}{2 \braket{\Psi|\Psi}}.
	\end{equation}
	We reduce this to an integral over the nodal surface by integrating over $z$.  Absorbing the Jacobian determinant  into $dA$ we obtain:
	\begin{equation} 
	\delta  E  =  -\frac{\int_{node}  d{\bf A}  \;
		g({\bf A})[\delta\Psi({\bf A},0)+\delta\Psi({\bf A},\eta)]}{2 \braket{\Psi|\Psi}} .
	\end{equation}
	From Eq.~\ref{eqn:bondaryConditions} we find that $\delta\Psi({\bf A},0)+\delta\Psi({\bf A},\eta) = g({\bf A}) \eta({\bf A})$ both when $\eta({\bf A}) > 0$ and when $\eta({\bf A}) < 0$. This yields 
	\begin{equation} 
	\delta  E =  \frac{ \int_{node} d{\bf A}  \;
		\eta({\bf A}) g^2({\bf A})}{2\braket{\Psi|\Psi}}  .
	\end{equation}
	By taking a derivative with respect to $\theta_i$ we derive our analytical expression	
		\begin{equation}  
	\frac{\partial E}{\partial\theta_i} = 
	\frac{\int_{node} d{\bf A} \:  |\nabla\Psi({\bf A})|^2 \partial_{\theta_i}\eta({\bf A})}{2\braket{\Psi|\Psi} }\:   .
	\label{eqn:EnergyDerivative}
	\end{equation}

	\subsection{Practical methods to estimate the energy gradient}
	
	Starting from Eq.~\ref{eqn:EnergyDerivative} we have derived three different equations for calculating the energy gradient from walker distributions, which we refer to as method A, B, and C. Since	DMC usually samples terms linear in $\Psi$, but  Eq.~\ref{eqn:EnergyDerivative} contains two terms bi-linear in $\Psi$, 
	an approximation used by methods A and B replaces the bi-linear terms with a mixed estimate of the true ground-state wavefunction and the trial function, i.e., 
	\begin{equation}  
	\label{eqn:Substitutions}
	\braket{\Psi|\Psi} \rightarrow \braket{\Psi_t|\Psi}, \:\:\:\:\:\:\:\:\:\: |\nabla \Psi|^2 \rightarrow \nabla \Psi_t \boldsymbol{\cdot} \nabla \Psi.
	\end{equation}
	Then, using
	\begin{equation}
	\frac{\partial \eta}{\partial \theta_i} = -\frac{1}{|\nabla \Psi_t|}  \frac{\partial \Psi_t}{\partial \theta_i},
	\end{equation}
	we approximate Eq.~\ref{eqn:EnergyDerivative} with
	\begin{equation}  
	\label{eqn:ApproximateEnergyDerivative}
	\frac{\partial E}{\partial\theta_i} 	\approx 
	-\frac{\int_{node} dA \:   |\nabla \Psi(A) | \partial_{\theta_i} \Psi_t(A) }{2\braket{\Psi_t|\Psi}} \:   .
	\end{equation}
	Unfortunately, this relies on the quality of $\Psi_t$, although this quality is improved with PGD. Method C does not make this approximation, thus in principle requires no Jastrow optimization.

	\subsection{Method A: the energy gradient from a standard walker distribution}

	Method A calculates Eq.~\ref{eqn:ApproximateEnergyDerivative} with a standard walker distribution equal to $\Psi_t \Psi$, which we think will make it easiest to implement into existing DMC code. It requires the evaluation of $\Psi_t$, its parameter derivative, and the parameter derivative of its $3N$ dimensional Laplacian. To get Eq.~\ref{eqn:ApproximateEnergyDerivative} in a form where this is possible, we change the nodal integral to a volume integral. Using the $z$ coordinate as defined in Sec.~\ref{section:Method} and Gauss' theorem, we transform Eq.~\ref{eqn:ApproximateEnergyDerivative} to
	\begin{equation}
	\label{eqn:SmallVolumeIntegral}
	\frac{\partial E}{\partial\theta_i} 	\approx 
	-\frac{1}{2\braket{\Psi_t|\Psi}}  \: \int_{node} d{\bf A} \: \lim_{\epsilon\to 0} \: \int_{-\epsilon}^{\epsilon} dz  \:  \frac{\partial \Psi_t}{\partial\theta_i} \nabla^2 \Psi \:,
	\end{equation}
	where we used the fact that $ |\nabla \Psi| = \tfrac{\partial \Psi}{\partial z} $ for $z>0$ and $\Psi=0$ for $z<0$.

	We replace $\nabla^2$ with $2({E-H})$ in Eq. \ref{eqn:SmallVolumeIntegral} in order to increase the bounds of the $z$ integral to all space, which is possible since $(E-H)\Psi=0$ everywhere except the nodes. We then integrate by parts to obtain
	\begin{equation}  
	\label{eqn:ApproximateEnergyDerivativeGuided}
	\frac{\partial E}{\partial\theta_i} 	\approx 
	\frac{\int_{vol} dR \: \: \Psi  ({H-E})  \partial_{\theta_i} \Psi_t  }{\int_{vol} dR \: \Psi_t \Psi } \:.
	\end{equation}
	Interestingly, one can also arrive at Eq. \ref{eqn:ApproximateEnergyDerivativeGuided} by taking the derivative of the mixed estimator $ \frac{\bra{\Psi} \footnotesize{ {H}}\ket{\Psi_t}}{\braket{\Psi | \Psi_t}}$ with $\Psi$ fixed. 
	
	Method A calculates the integrals of Eq.~\ref{eqn:ApproximateEnergyDerivativeGuided}  from positions $R_k$ and weights $w_k$ of a walker distribution equal to $\Psi_t \Psi$ with the equation
	\begin{equation}  
	\label{eqn:master3}
	\frac{\partial  E}{\partial\theta_i}   \approx  \frac{\sum_{k}
		w_k \Psi_t^{-1}(R_k)      ({H-E})   \partial_{\theta_i}
		\Psi_t(R_k)  }{\sum_{k} w_k } \: .
	\end{equation}

	\subsection{Method B: the energy gradient from a nodal walker distribution}
	
	Method      B requires the evaluation of $\Psi_t$ and its parameter derivative. It      samples      $|\nabla       \Psi|$      of
	Eq.~\ref{eqn:ApproximateEnergyDerivative}   directly  from   a
	walker distribution on the nodal surface, which we generate by
	recording  walkers that cross the nodal-surface, and are thus deleted. We note that shifting crossed  walker positions  closer to  the nodes using Newton's method noticeably improves performance. Unlike those of method A,
	the pre-crossed walkers cannot  be  guided  by  $\Psi_t$. This is partly  because no  walkers will  cross the  nodes of $\Psi_t$ in the zero time step limit. Instead we guide them with a node-less function $\Psi_g$. 
	
	We wish to relate $|\nabla\Psi|$  at the nodal-surface  to the  number of
	walkers that cross  the nodal-surface per area  $dF/dA$ during time
	$\Delta \tau$. Let us assume a constant fixed-node walker distribution  equal  to  $\Psi_g\Psi$. If we suddenly remove the nodal-surface, there will be an initial increase of walker population resulting from walkers that would have crossed the nodal-surface where it present. Therefore the resulting population increase
	per nodal area  to first order in time is  $dF/dA$. Let us
	make $\Delta \tau$ small enough that $\Psi$ does not change at
	a  distance  $\epsilon$ from  the  node;  then, the  population
	increase per nodal area from the released nodes is
	\begin{equation}
	\frac{dF(A)}{dA}     =    \Psi_g({\bf R}_A)\int_{-\epsilon}^{\epsilon}
	\int_0^{\Delta \tau} dz d\tau \frac{d\Psi({\bf A},z)}{d\tau},
	\end{equation}
	where ${\bf R}_A=({\bf A},0)$. We can rewrite the above as 
	\begin{equation}
	\label{eqn:smallVolumeTimeIntegral}
	\frac{dF(A)}{dA}     =    \Psi_g({\bf R}_A)\int_{-\epsilon}^{\epsilon}
	\int_0^{\Delta \tau} dz d\tau (-{H}) \Psi({\bf A},z).
	\end{equation}
	
	We take the $\epsilon \to 0$  limit of Eq. \ref{eqn:smallVolumeTimeIntegral}, which leaves only a kinetic contribution from the
	discontinuity of $\nabla \Psi$:
	\begin{equation}
	\frac{dF(A)}{dA}     =    \Psi_g({\bf R}_A)\int_{-\epsilon}^{\epsilon}
	\int_0^{\Delta \tau} dz  d\tau \tfrac{1}{2} \nabla^2 \Psi({\bf A},z).
	\end{equation}
	Then, using
	\begin{gather} 
	\nabla^2\Psi({\bf A},\epsilon)= \partial_z^2\Psi({\bf A},\epsilon) \notag \\ 
	|\nabla \Psi ({\bf A},\epsilon)| = \partial_z \Psi ({\bf A},\epsilon)\notag \\  | \nabla \Psi({\bf A},-\epsilon) |=0,
	\end{gather} 
	we write
	\begin{equation}
	\label{eqn:Flux2}
	\frac{dF(A)}{dA}                                             =
	\tfrac{1}{2}\Delta\tau\Psi_g({\bf R}_A)|\nabla\Psi({\bf R}_A)| \: .
	\end{equation}
	
	We can now use Eq.~\ref{eqn:Flux2}  to sample the $|\nabla\Psi({\bf R}_A)|$ term of
	Eq.~\ref{eqn:ApproximateEnergyDerivative} which yields the
	equation
	\begin{equation}  
	\frac{\partial  E}{\partial\theta_i} \approx  -\frac{ \sum_{j} w_j \Psi^{-1}_g(R_j)
		\; \partial_{\theta_i} \Psi_t(R_j)
	}{\Delta \tau \braket{\Psi_t|\Psi}},
	\end{equation} 
	where $R_j$ and $w_j$ are, respectively, the positions and weights of walkers that
	crossed   the  nodes   during a the period $\Delta\tau$.   To  sample   the
	denominator, we simply use
	\begin{equation} 
	\Delta \tau \braket{\Psi_t|\Psi} = \int^{\Delta \tau}_{0}d\tau
	\sum_{k} \frac{\Psi_t(R_k,\tau)}{\Psi_g(R_k,\tau)} w_k(\tau),
	\end{equation}
	with $R_k$ and $w_k$ the  positions and weights of all walkers
	as a function of time $\tau$. 
	
	Method B therefore calculates the energy gradient with
	\begin{equation}  
	\label{eqn:master}
	\frac{\partial  E}{\partial\theta_i} \approx  -\frac{ \sum_{j}
		w_j   \Psi_g^{-1}(R_j) \partial_{\theta_i}\Psi_t(R_j) 
	}{    \int^{\Delta    \tau}_{0}d\tau
		\sum_{k} w_k \Psi_g^{-1} (R_k)\Psi_t(R_k) }.
	\end{equation} 
	
	\subsection{Method C: the exact energy gradient}
	
	Method C does not use the approximation of
	Eq.~\ref{eqn:Substitutions}. It samples one factor of $|\nabla \Psi|$ in Eq.~\ref{eqn:EnergyDerivative} with the same nodal walker distribution used by method B, and samples the other factor of $|\nabla \Psi|$ using forward walking. Just as with method B, method C requires the evaluation of $\Psi_t$ and its parameter derivative.
	
	We start by writing the $\Psi$ projector as a time evolution operator
	\begin{equation}  
	\label{eqn:projector1}
	\frac{\ket{\Psi}\bra{\Psi}}{\braket{\Psi | \Psi}} = \lim_{\tau
		\to \infty} \exp\big[\tau({E\text{-}H})\big]
	\end{equation} 
	%	We rewrite the time evolution operator  with the       modified
	%	Hamiltonian $\widetilde{{H}}  \equiv  {\Psi}_t  {H}
	%	{\Psi}_t^{\text{-}1}  $, which is used for a walker guided by $\Psi_t$
	%	\begin{equation}  
	%	\label{eqn:projector2}
	%	\exp\big[\tau({E-H})\big] =  {\Psi}_t^{\text{-}1}
	%	\exp\big[\tau({E-}\widetilde{{H}})\big] {\Psi}_t .
	%	\end{equation} 
	%   We then insert this projector into the following equation
	%    	\begin{equation}  
	%	\label{eqn:projector3}
	%	\bra{\Psi_t} \frac{\ket{\Psi}\bra{\Psi}}{\braket{\Psi | \Psi}}  \Psi_t^{-1} \ket{\bf R},
	%	\end{equation} 
	and contract the left side of Eq.~\ref{eqn:projector1} with $\bra{\Psi_t}$ and the right with $  | J \rangle_{\bf R}  $ where
\begin{eqnarray}
  | J \rangle_{\bf R} \equiv   {1 \over
    {\Psi_t({\bf R})}} | {\bf R} \rangle,
  \end{eqnarray}
resulting in
	\begin{equation} 
	\label{}
	%	\begin{split}
	\frac{\Psi({\bf R}) \braket{\Psi_t |  \Psi} }{ \Psi_t({\bf R})  \braket{\Psi | \Psi}}  =    \\ \lim_{  \tau \to
		\infty       }       \bra{\Psi_t}       
	\exp\big[\tau(E\text{-}H)\big]  |J \rangle_{{\bf R}}.
	%	\end{split}
	\end{equation}
	We recognize that
	\begin{equation} 
	\bra{\Psi_t} \exp\big[\tau(E\text{-}H)\big]  | J \rangle_{\bf R} =
	\int d{\bf R}' \widetilde{G}({\bf R}',{\bf R},\tau) ,
	\end{equation}
	where    $\widetilde{G}$, a  Green's function, is the evolution operator of $\Psi_t \Psi$. Because $ \Psi({\bf R}) / \Psi_t({\bf R}) $ approaches $ |\nabla\Psi({\bf R}_A)|/|\nabla\Psi_t({\bf R}_A)| $ as $\bf R$ approaches the nodal surface (since $\nabla =\partial_z$ at $z \to 0^+$), at the nodal-surface we have
	\begin{equation} 
	\label{eqn:g_w3}
	|\nabla\Psi({\bf R}_A)|  = |\nabla\Psi_t({\bf R}_A)|  \frac{\braket{\Psi |
			\Psi}}{\braket{\Psi_t |  \Psi}} \lim_{  \tau \to  \infty }
	\int d{\bf R}' \widetilde{G}({\bf R}',{\bf R}_A,\tau) .
	\end{equation} 
	
	We now substitute  one  $|\nabla\Psi|$  term of
	Eq.~\ref{eqn:EnergyDerivative}  with   Eq.~\ref{eqn:g_w3}  and
	again   use  $\frac{\partial   \eta}{\partial  \theta_i}   =
	-\frac{1}{|\nabla  \Psi_t|}   \frac{\partial  \Psi_t}{\partial
		\theta_i} $ to get
	\begin{equation}  
	\label{eqn:CorrectedEnergyDerivative}
	\begin{split}
	\frac{\partial              E}{\partial\theta_i}             = -
	\frac{1}{2\braket{\Psi_t|\Psi}}\:     \int_{node}    d{\bf A}     \:
	\frac{\partial \Psi_t}{\partial\theta_i}  \: |\nabla \Psi({\bf R}_A)
	| \\   \times  \lim_{   \tau  \to   \infty  }   \int  d{\bf R}'   \:
	\widetilde{G}({\bf R}',{\bf R}_A,\tau) .
	\end{split}
	\end{equation}
	
	Eq.~\ref{eqn:CorrectedEnergyDerivative}   is    identical   to
	Eq.~\ref{eqn:ApproximateEnergyDerivative} except the added factor $\lim_{  \tau  \to  \infty }  \int  d{\bf R}'
	\widetilde{G}({\bf R}',{\bf R}_A,\tau)$. We sample  this factor with the value $\xi$ we define as the final weight of a walker that had an initial weight of one and an initial position of  ${\bf R}_A$, and was then propagated for time  $\tau$  with
	$\Psi_t$ as the guiding function, that is
	\begin{equation}
	\label{eqn:g_w4}
	\big{<}\xi({\bf R}_A,\tau)\big{>}         =         \int         d{\bf R}'
	\widetilde{G}({\bf R}',{\bf R}_A,\tau).
	\end{equation} 
	Instead of taking the $
	\tau \to \infty $ limit, we  cut the propagation time short at
	$\tau_c$. The value we choose should be long  enough for $\big{<}\xi({\bf R}_A,\tau)\big{>}$ to become roughly constant.
	
	Placing a walker guided by $\Psi_t$ at the  nodes can  be problematic since the
	local energy and raw drift velocity ${\bf V}=\nabla \ln{\Psi_t}$ are
	divergent  at  the  nodes when  $\Psi_t$ is used as the guiding function. When calculating $\xi$, we 
	suppress the effects of both divergences in the standard way, where we  modify the drift velocity
	and  weight     increase  by  multiplying them with    $\tfrac{-1     +
		\sqrt{1+2V^2 \delta\tau}}{V^2 \delta\tau}$, where $\delta\tau$ is the time step. To insure walkers move towards the positive side of the node, we also multiply the drift velocity
	by  $\sign(\Psi_t)$. 
	
	Method C  calculates Eq.~\ref{eqn:CorrectedEnergyDerivative}
	with node crossing walkers following the same logic we used for method B and Eq. \ref{eqn:ApproximateEnergyDerivative}, but
	with the extra factor $\xi$.  	We multiply
	Eq.~\ref{eqn:master} with $\xi(R_j,\tau_c)$ to obtain our equation for method C,
	\begin{equation}  
	\label{eqn:master4}
	\frac{\partial  E}{\partial\theta_i} \approx  -\frac{ \sum_{j}
		w_j  \xi(R_j,\tau_c) \Psi_g^{-1}(R_j)  \partial_{\theta_i}\Psi_t(R_j) 
	}{    \int^{\Delta    \tau}_{0}d\tau
		\sum_{k} w_k \Psi_g^{-1} (R_k)\Psi_t(R_k) }.
	\end{equation}

	\section{Implementation}
	\label{section:Implementation}
	
	To test our method, we  developed a DMC  code  with  walker
	propagation that for the  most  part follows  the prescription  of
	Umrigar  et   al.\cite{UmrigarEtAl}.
	In addition, we introduced our PGD iteration method in the code.
	The main steps are outlined next. We will apply our technique on three systems, atomic  Be, Li\textsubscript{2},
	and atomic Ne.
	
	\subsection{Trial and guiding function}

	The $\Psi_t$ used  for all of our calculations were of
	the well-known Slater-Jastrow    form
	\begin{eqnarray}
	\Psi_t     = J \Psi_S, \hskip 0.2 in \Psi_S = D^{\uparrow}D^{\downarrow}, 
	\end{eqnarray}
	where    $D^{\uparrow}$    and
	$D^{\downarrow}$  are  spin-up  and  spin-down  Slater
	determinants of single-particle orbitals $\phi_a(r)$ described in the
	next paragraph. $J$ is a Jastrow function
	\begin{equation}
	J= \prod_{i<j} f_{ij}, \hskip 0.2 in  
	f_{ij}= e^{-{{a_{ij} r_{ij}} \over {1 + b r_{ij}}}}
	\end{equation}
	where $r_{ij}$ is the electron-electron distance and $a_{ij}$ is equal to 1/4 (or 1/2) when both $i$ and $j$ correspond
	to electrons of parallel (or anti-parallel) spin projections\cite{UmrigarEtAl}.

	The single-particle orbitals $\phi$ of the Slater determinants
	are made of a basis of Slater functions, taking the form
	\begin{eqnarray}  
	\label{eqn:Sorbital}
	\phi_a(r) = \sum^{N_{basis}}_{b=1} C_{ab} r_b^{n_b-1}e^{-\xi_b
		r_b}Y_{l_b m_b}(\hat{r}_b),
	\end{eqnarray}
	where $r_b = |{\bf r-R_b}|$, and
	$n_b$, $l_b$,  and $m_b$ are quantum  numbers characterizing the $b$ basis state
	which is centered at the atomic nuclear position ${\bf R}_b$. The parameters we optimize are the
	$C_{ab}$  and $\xi_{b}$,  with $\xi_{b}$  being shared  by all
	orbitals.
	
	Since $J$ is  node-less, and, thus, does not  affect the energy,
	our method cannot  improve it. We therefore do  not include it
	during  PGD since  we have  no a priori  knowledge of  its
	parameters. Although there is a reason  to expect that a quality $J$
	will  improve the methods  A and  B, since  they rely  on a  $\Psi
	\rightarrow \Psi_t$  approximation, we  do not notice  such an
	improvement.  Method  C  on  the  other  hand  makes  no  such
	approximation, so  in principle, it  does not  require an optimized $J$.  We only
	include $J$ when evaluating the energy after optimization.
	
	Our walker distributions for methods B and C use an electron-nuclear
	Jastrow function as a guiding function $\Psi_g$.  It has the form
	\begin{equation}
	\Psi_g =
	\prod_{i,k} \exp\Bigg{(}-\frac{Z_k  |{\bf R_k}-{\bf r}_i|  }{1+|{\bf R_k}-{\bf r}_i|}\Bigg{)},
	%\nonumber
	\end{equation}
	with ${\bf R_k}$ the nuclear coordinate, $Z_k$ the nuclear charge,   and  ${\bf r}_{i}$  the electron coordinate.  We  choose this  $\Psi_g$ instead of $\Psi_g=1$  to reduce
	the statistical, time-step, and  population growth errors. This
	$\Psi_g$ also results  in many more walkers  crossing the node
	per time, and the electrons are far less likely to ionize. Ionization can still be a problem with poor nodes, when this is the case, we surround the system with a potential barrier.

	\subsection{Projection on the cusp-condition satisfying parameter space }

	The electron-nuclear cusp-condition\cite{Kato,Steiner} is given by
	\begin{equation}
	S_k \equiv -\frac{1}{2} \frac{d \ln{\overline{\Psi_t^2}}}{dr} \Big\rvert_{r=0} \;,\;\;\;\; S_k= Z_k,
	\end{equation}	
	where $r_k$ is the radial coordinate from nucleus $k$, $Z_k$ is the nuclear charge, and $\overline{\Psi_t^2}$ is defined as the angular average of $\Psi_t^2$ about $r_k=0$. It is a necessary requirement to avoid  a large divergence of local
	energy near  the nucleus, which gives rise to a large increase  of all
	errors besides the fixed node  error. Since the cusps $S_k$ of our
	$\Psi_t$  depend   on $\Psi_S$ and its varied parameters  $\theta_i$, we  project
	$\theta_i$ back  to the manifold of the cusp  condition after
	each iteration of gradient descent.

	The cusp condition of $\Psi_S$  is satisfied when the cusps $S_{ka}$ of all
	single-particle orbitals $\phi_a$  satisfy $S_{ka} = Z_k$. 	We assume the shift in parameters is small enough (due to small $a_i$) that a linear approximation of the cusp $\widetilde{S}_{ka}$ can be made at  the pre-projected
	parameters $\theta_{i0}$:
	\begin{eqnarray}
	\widetilde{S}_{ka}(\theta_i)      \equiv      S_{ka}(\theta_{i0})     +      \sum_i
	(\theta_i-\theta_{i0}) \frac{ \partial	S_{ka}(\theta_{i0})} {\partial\theta_i}.
	\end{eqnarray}
	$\theta_i$ is then shifted to the closest point satisfying
	\begin{equation}
	\widetilde{S}_{ka}(\theta_i)= \widetilde{S}_{ka}(\theta_{i0}) + c(Z_k -\widetilde{S}_{ka}(\theta_{i0}) )	,
	\end{equation}		   
	where $c$ is added for stability and is between zero and one (we used 0.5). We repeat this until $S_{ka} - Z_k$ is within a threshold. 
	
	\iffalse	
	
	\item       Our  propagator is  independent of  the nuclei
	positions. Also,  when propagating  with a  node-less $\Psi_g$
	for methods B and C, we  do not use an accept-reject walk and
	we decrease the time-step size of walkers as they  approach the
	node.
	\fi

	\subsection{Details of the PGD iterations}
	%        \begin{itemize}
	
	%      \item
	After each PGD iteration, we update the  energy and effective time-step\cite{doi:10.1063/1.443766} with   data  of that  iteration.  We use  the  mixed
	estimate
	\begin{equation}
	\hskip 0.2 in           E =   \frac{\bra{\Psi_t}       \footnotesize{
			{H}}\ket{\Psi}}{\braket{\Psi_t | \Psi}},
	\nonumber
	\end{equation}
	for method A,
	and for methods B and C we use the growth estimate
	\begin{equation}
	\hskip 0.2 in   E =  \Delta\tau^{-1} \ln{{ \Sigma(0)}\over {\Sigma(\Delta\tau)}}, 
	\nonumber
	\end{equation}
	where $\Sigma(\tau)$ is the sum of the weights of all the walkers at imaginary time $\tau$,  ignoring weight normalization.

	%\item
	To   calculate  the   gradient,   we   propagate  the   walker
	distribution $f$ until the number of samples (walker positions
	and weights)  reaches a  threshold. Samples  for method  A are
	taken every  time-step, and  samples for methods B-C  are taken
	from  all node-crossings walkers. After  an iteration of projected gradient
	descent, we propagate $f$ for an extra time (around 0.1 Ha$^{-1}$)
	to  adjust it to the  new $\Psi_t$ before taking  samples for
	the  next gradient.  This  is  necessary to  do  for method  A
	because after a shift of the node,  $f$ does not go to zero at
	the    nodes    fast    enough     for    the expectation of the sum of Eq.~\ref{eqn:master3} to  be finite,
	due     to      the     $\Psi_t^{-1}$      factor.
	For method A, requiring   $f$   to   go   to   zero
	sufficiently fast  also requires us to use an accept-reject
	step.

	%\item
	The  gradient  has  stochastic  error that  prevents PGD of the
	parameters $\theta_i$  from fully settling to an accurate minimum, where the signal to noise ratio of the gradient diverge. Instead, $\theta_i$ continues  to fluctuate
	around the minimum after some PGD iterations. We  argue in  Appendix~\ref{section:Fluctuations} that to first order in the $a_i$  of  Eq.~\ref{eqn:GradientDescent},
	the variance of fluctuations is proportional
	to both the  variance of the gradient error and 
	$a_i$.  Thus we can suppress the fluctuations either through smaller $a_i$ and/or by reducing the statistical error of energy derivative with more samples.
	% In  addition, some level  of stochastic error may  be desirable and is often  added to  the gradient in machine learning applications.

	%\item
	An  important  choice to make is the value of  $a_i$. The optimal  value 
	depends on  many factors,  such as the parameters $\theta_i$, the  form of  $\Psi_t$, the
	number  of PGD  iterations, and  the stochastic gradient error. Generally,
	decreasing  $a_i$ requires  more  PGD iterations  to
	reach a minimum, and may exacerbate the problem of getting stuck in local
	minima. While  increasing $a_i$ increases
	fluctuations from the gradient error and may cause overshooting of $\theta_i$ from its optimum value. 
	
	%\item
	The most  basic gradient  descent scheme  uses a  common value of
	$a_i$  for   all  parameters   and  iterations.    We
	experiment  with this
	and  with the two  adaptive  gradient  descent  algorithms  ADAM\cite{ADAM} and
	RMSprop\cite{RMSprop}.  The  adaptive   algorithms  make
	$a_i$ inversely proportional to a trailing root square mean of past  gradient components,
	and are described more in Appendix~\ref{section:Algorithms}. The  $a_i$ of all our algorithms have a common factor which we label the
	learning rate $\alpha$, which we determine empirically. We sometimes decrease this value with PGD iterations to suppress  fluctuations of  $\theta_i$. In Appendix~\ref{section:EffectiveAlpha} we examine the effective rang of $\alpha$ for different gradient descent algorithms and systems.

	%,  and   ADAGrad\cite{Adagrad}
	
	%and  others, including  an
	%$\alpha$ that decreases reciprocally with PGD iterations,

	%\end{itemize}	

	\section{Results}
	\label{section:Results}
	We tested our methods for three systems, atomic  Be, Li$_2$, and atomic  Ne, and we compared our results to those obtained  by the
	DMC calculation of Umrigar  et  al.\cite{UmrigarEtAl} where the fermion nodes were optimized using VMC. %The trial functions of Be, Li$_2$, and Ne had respective parameter numbers of 9, 64, and 40. However, due to symmetry and the cusp condition, the DMC energy of Be, Li$_2$, and Ne in principle depended on only 3, 4, and 9 parameters respectively.  

	\subsection{Derivative of the energy calculation}
	
	\begin{figure}[!htb]
		\centering
		
		\includegraphics[width=1.\linewidth]{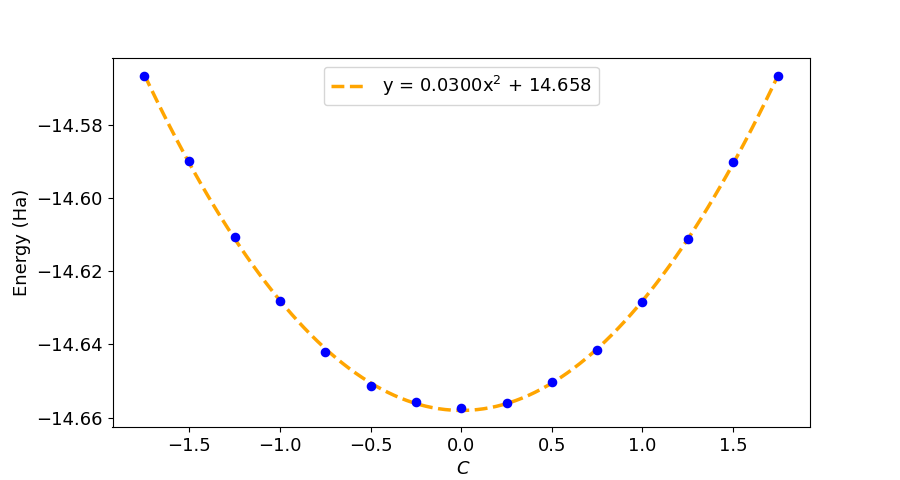} % BeEnergy.png}
		
		\includegraphics[width=1\linewidth]{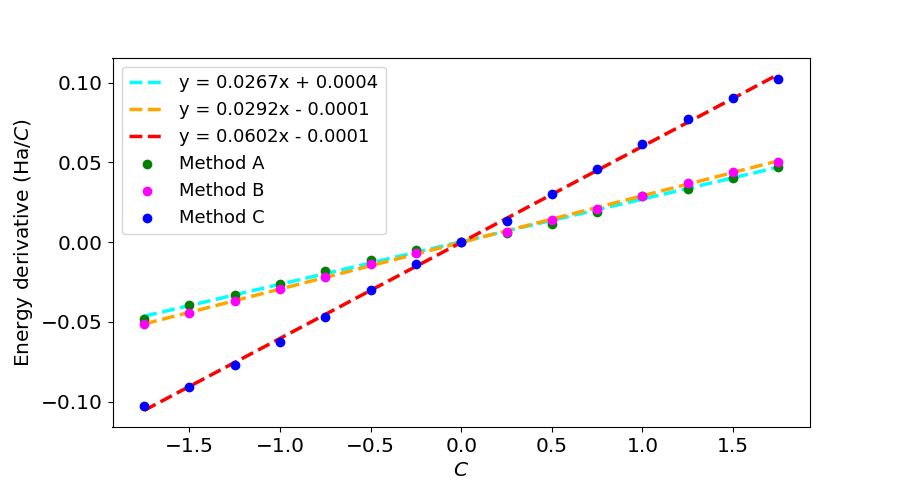} % EnergyDerivative
		
		\includegraphics[width=1\linewidth]{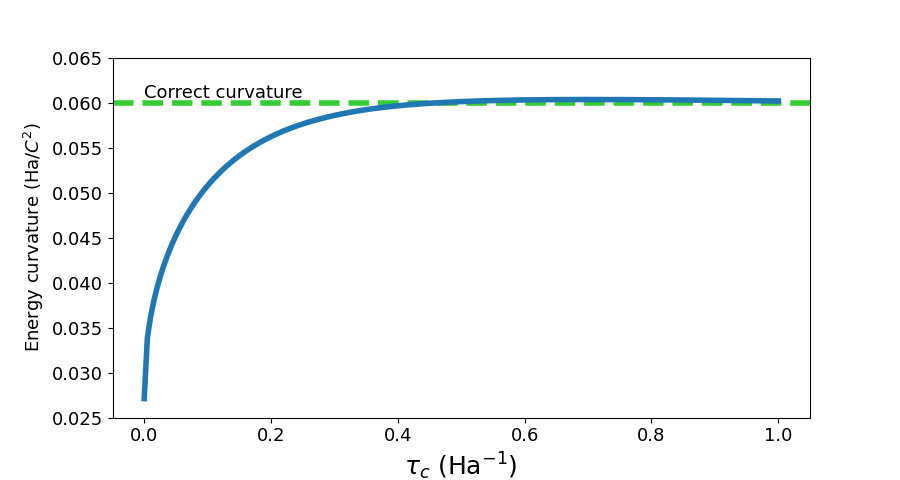} % MethodC
		
		\caption{We compare the DMC energy (top) of a Be trial function for a varied parameter   with the energy derivative with respect to said parameter (center) that is calculated with methods A-C. Data points are shown as dots with an associated best fit line or parabola. The  parabola curvature is within  1\% of the slope of method C, which uses $\tau_c$  = 1  (Ha$^{-1}$). We examine how the slope of method C (bottom) depends on $\tau_c$. }
		\label{fig:plot1}
	\end{figure}

	We present tests of the energy gradient calculation methods A, B, and C in this subsection. They examine one parameter of a basis we added to the simple Be trial-function that is fully described in reference \cite{UmrigarEtAl}. The basis has quantum numbers $n=2$, $l=1$, and $m=0$, and the parameter is the non-normalized coefficient of the basis. We varied this parameter and obtained its partial first and second energy derivatives by means of a fixed node DMC calculation, which is shown in the upper panel of Fig.~\ref{fig:plot1}. 
	As can be noticed, the energy obtained from the fixed node DMC is well approximated with a parabola. We wish to compare the derivative of the parabola with the energy derivatives obtained by our three methods which are shown in the center panel of Fig.~\ref{fig:plot1} and are well approximated with lines intersecting the origin.
	
	The energy derivative obtained by method C agrees with that from the parabola to within 1\%, while the energy derivative obtained using  methods A and B is proportional to that from the parabola but off by a factor of 2, which we attribute to the approximation of Eq.~\ref{eqn:Substitutions}. We expect that when each of the components of the energy derivative
        has the correct sign it results in shifting the parameters towards the direction where the energy is lower.
        In addition, many
	gradient  descent  algorithms   divide each gradient component by a trailing root-mean-square  of past gradient components, thus,  the exact value of the slope is not needed by such algorithms.
	As a result methods A and B  may     still be useful.

	The bottom panel of Fig.  \ref{fig:plot1} shows how the energy derivative obtained with method C depends on the cut off time $\tau_c$. Our choice of $\tau_c$ when optimizing with method C is based on this dependence.

	\begin{figure}[!htbp]
		\centering
		\includegraphics[width=1\linewidth]{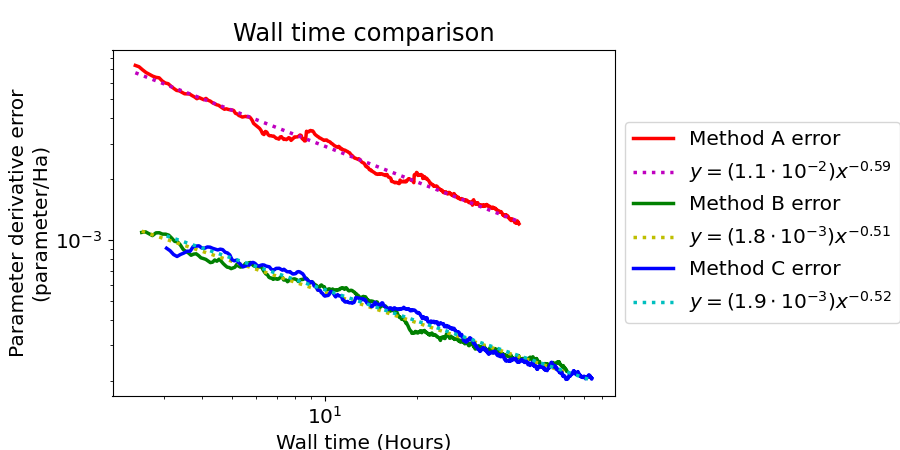} % GradErrors.png
		\caption{We compare the efficiency of methods A, B, and C by plotting their energy derivative error for the parameter of Fig. \ref{fig:plot1} as wall-time (sample size) is increased, using a common time step. Best-fit (dotted-lines) power laws are roughly proportional to the inverse square root of sample size. Methods B and C require similar wall-time to achieve the same error, while method A requires on the order of fifty times more.}
		\label{fig:plot8}
	\end{figure}
	
	We compare the efficiency of the three methods in Fig. \ref{fig:plot8} by plotting the error of the energy derivative versus wall-time, which is proportional to the sample size. In this Figure we show the best-fit power laws which, as expected, are approximately proportional to the inverse square root of sample size. Method B and C seem to require similar wall-time to reach a given error level when using $\tau_c=0.25$ Ha$^{-1}$. Method A on the other hand is considerably more expensive, though a fair comparison is difficult partly because method A may not require the same time step, and the error depends both on the associate parameter and on $\Psi_t$.

	\subsection{Projected gradient descent evolution}
	
	%	\FloatBarrier
	
	\begin{figure}[!htb]
		\centering
		\includegraphics[width=1\linewidth]{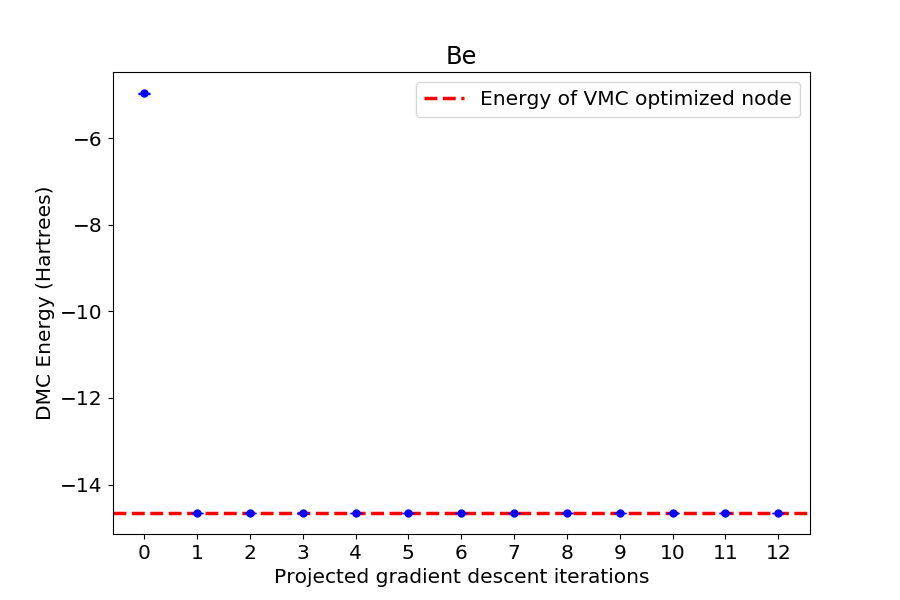} % Be_EnergyVsIterations
		\centering
		\includegraphics[width=1\linewidth]{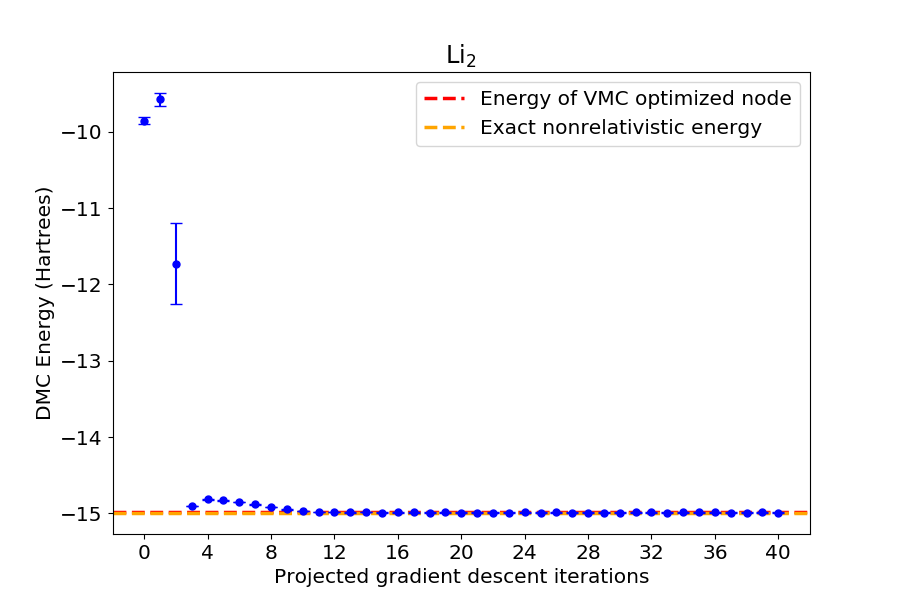} % Li$_2$_EnergyVsIterations  Fig3b
		\centering
		\includegraphics[width=1\linewidth]{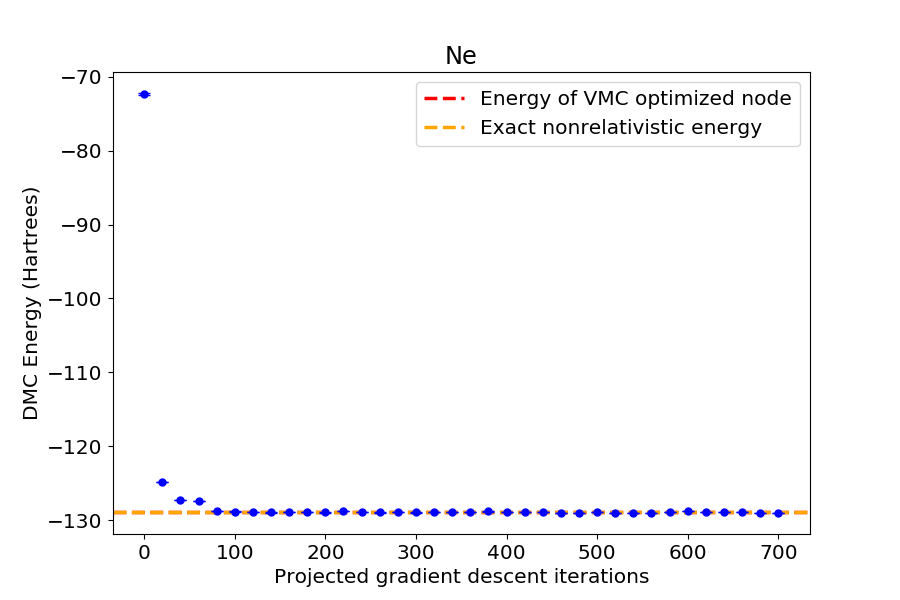} % Ne_EnergyVsIterations  Fig3c
		\centering
		\caption{ Projected gradient descent of initially random parameters. See text for detailed explanation. }%  ADAM was used for Li$_2$ and Ne, while Max was used for Be. The parameters used for each run displayed in the above table. }
		\label{fig:plot6}
	\end{figure}

	\begin{figure}[!htb]
		\centering
		\includegraphics[width=1\linewidth]{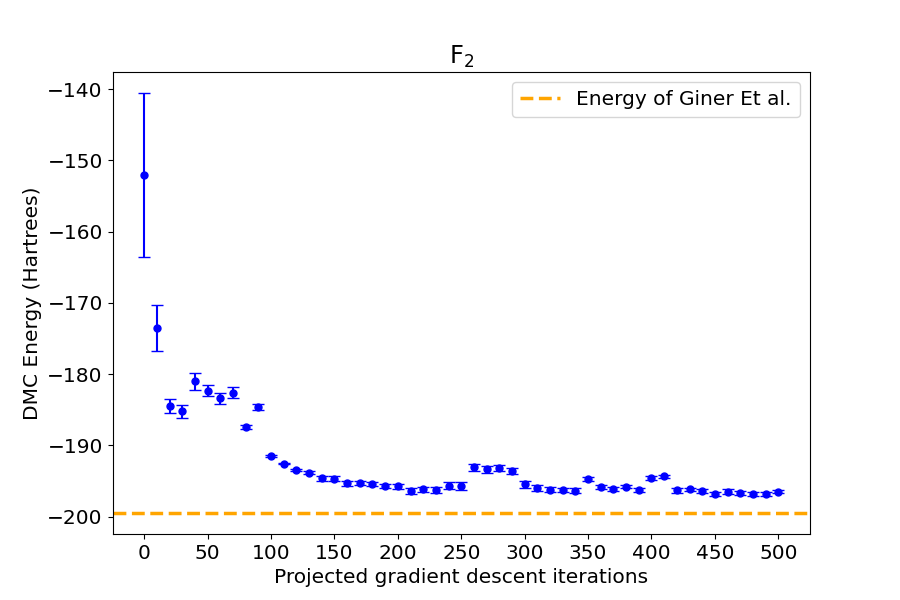} % Ne_EnergyVsIterations  Fig3c
		\centering
		\caption{  See text for detailed explanation. }%  ADAM was used for Li$_2$ and Ne, while Max was used for Be. The parameters used for each run displayed in the above table. }
		\label{fig:F2}
	\end{figure}

	In order to demonstrate the utility of the method, we start from random values for the parameters $C$ and $\xi$ of all-electron single-determinant wavefunctions of Be, Li$_2$, Ne,  and we
	perform PGD iterations with our cusp condition projection. We present a history of the energy of their nodes in Fig. \ref{fig:plot6}. One can see the DMC energies rapidly decrease with iteration to the energy of the VMC optimized nodal surface obtained by Umrigar et. al.\cite{UmrigarEtAl} (red dotted lines). Ne  continues to fluctuate after initial decrease, which we attribute to statistical error of the energy gradient and an excessively large decent rate ($a_i$ term of Eq.~\ref{eqn:GradientDescent}). ADAM was used for all examples, and we decreased the descent rate ($\alpha$ parameter of Eq.~\ref{eqn:ADAM}) with iteration number. Method C was used for all except  Be, where method A was used, demonstrating the utility of the approximation of Eq.~\ref{eqn:ApproximateEnergyDerivative}.

	We chose these three systems because a full description of their VMC optimized wavefunctions was present in the literature\cite{UmrigarEtAl}, allowing us to directly compare our PGD optimization using wavefunctions of the same form.  There are DMC calculations for larger atoms and many atom systems;
	however,         many of the
	atomic electrons (the inner) are absent in most of these calculations because they are using pseudopotentials and, therefore, do have to implement the
	cusp condition. Nevertheless, we have applied PGD to a somewhat larger system of an all electron single determinant of F$_2$, shown in Fig.~\ref{fig:F2}. Unfortunately we do not have a  wavefunction of the same form optimized by other means for comparison. However, a DMC calculation of F$_2$ was done by Giner {\it et al}\cite{F2DMC}, who used a configuration interaction optimized wavefunction consisting of $10^5$ determinants. They  achieved an energy of -199.2977(1) Ha, while our lowest PGD iteration was -196.7(3) Ha. This discrepancy can partly be attributed to the fact that our wavefunction had just one determinant (a current limitation of our code), and a single particle basis of just 20 Slater type orbitals. The decrease in error of the F$_2$ energy with iteration number suggests that PGD also indirectly reduces local energy variance.

	\subsection{Energy upper-bound estimation}
	The set $\{E_1,E_2,...,E_n,...\}$ formed by the values of the energy $E_i$ at  $i^{th}$ step of the PGD method shown in Fig.~\ref{fig:plot6} is to be taken as a set of energy upper bounds.
	As already discussed the variations from step to step are due to the fact that the sign of the energy derivative with respect to the PGD parameter $\alpha$ is sampled and, thus, its choice
	has some random element. For every such choice, the DMC evolution can be carried out with controlled level of error because the nodal surface is fixed.
	Therefore, the best energy upper-bound ${\cal E}$ corresponds to the lowest value of the energy in the set, i.e., ${\cal E} = \min\{E_1,E_2,...,E_n,...\}$.
	The last iteration energy obtained for the three systems studied in this work are compared with the best DMC energy obtained with nodes determined by a VMC optimization\cite{UmrigarEtAl}
	in Table~\ref{table:2}.
	
	The present calculation suggests that the nodes obtained with VMC optimization are not far from the optimum at least within the error of the present
	calculation. Therefore, the small discrepancy with the exact nonrelativistic values could be attributed to the fact that the form of the one-body factor is limited
	with the space of a single Slater determinant and the lack of spin-spin correlations  because the one-body factor is taken to be
	a product of the form $D^{\uparrow}D^{\downarrow}$.

	\begin{table}[!h]  
		\begin{tabular}{ |c|c|c|c|c| }
			\hline
			System &  Last PGD iteration &  VMC optimized nodes  & Exact energy \\
			\hline
			Be &  -14.6567(5)  & -14.6571(1)  & -14.66736 \cite{BeLiEnergy}  \\          
			\hline
			Li$_2$  & -14.989(1) & -14.9898(1) & -14.9954 \cite{BeLiEnergy} \\
			\hline
			Ne & -128.92(1) & -128.919(3) & -128.939 \cite{NeEnergy} \\            
			\hline
		\end{tabular}
		\caption{DMC energy comparison between the last PGD iteration nodes with VMC optimized nodes. Exact nonrelativistic values are present. }
		\label{table:2}
	\end{table}			
	
	\FloatBarrier
	
	\section{Discussion}
	\label{section:Discussion}

	We are not yet comfortable claiming one of methods A, B, or C is ideal, and the best choice may depend on circumstance. For example, although method A was least efficient in our test, it may not require as small of a time step, and we think it would be the simplest to implement into existing code due to its use of a standard walker distribution. Although the efficiency of method C was comparable to that of method B in our test, it requires additional wall-time to calculate $\xi$, which also introduces the additional parameter $\tau_c$ which we do not yet have a good procedure for determining. 
	
	Nevertheless, method C performed the best in our tests and also has the advantage of not relying on the quality of $\Psi_t$. We recommend pairing method C with ADAM.  In our tests we were able to optimize the location of the nodes  to roughly the same energy as with the VMC
	optimized nodes using any of the methods and without using Jastrow factors in our guiding function.

	Without the requirement of an optimized Jastrow function, our method is independent of prerequisite optimization by VMC. Nevertheless, simultaneously optimizing the Jastrow and Slater parameters may be an fruitful avenue, as a quality Jastrow function reduces many sources of error (except for the fixed node error) and should improve the approximation of methods A and B. This could be done with the standard VMC method, or possibly with DMC and gradient descent, which require a gradient different than one of the DMC energy, perhaps one of the
	local energy variance.

	An important advantage of PGD optimization is that the use of DMC propagation produces correct correlations in the fixed-node wavefunction, whereas VMC optimizes the nodal surface within a limited Jastrow form, which captures only pairwise correlations. Thus, the exact path to optimization due to the interplay between the adjustment of these correlations
	as the position of the nodal surface changes and vice versa was not fully accounted in previous DMC studies.
	
	We would like to give an example of a system which demonstrates the significance of the present work. Variational calculations of liquid $^3$He, where
	just a Jastrow-Slater wavefunction is used, yield\cite{PhysRevB.28.3770} an un-physical result that the fully polarized liquid is of lower energy at its equilibrium density than
	the unpolarized liquid
	at its equilibrium density. Adding three-body correlation factors in the wave-function does not change the above conclusion. It is when
	one includes backflow correlations, which are state-dependent and modify the nodal surface of the variational wavefunction, that one
	finds\cite{PhysRevB.28.3770} that the unpolarized state of liquid $^3$He is energetically favorable. 
	
	Therefore, let us pretend that we do not know the fact that the
	naturally occurring liquid $^3$He is spin-unpolarized;  we might then be naive and 
	use a Slater-Jastrow variational calculation to determine the optimum polarization.
	The nodes that would be determined by such a VMC optimization
	would be those that correspond to a polarized liquid. A subsequent fixed node DMC calculation with these nodes is not going to change this result, and we would reach an unphysical conclusion.
	The PGD method described in the present paper, however, should find that the unpolarized state is the ground state. 
	%It may even be possible to find the correct polarization in one run by slightly modifying the Hamiltonian to allow the diffusion of walkers to different spin configurations.
	
	Another advantage of PGD is that it requires only one gradient calculation to shift all the parameters in the correct direction. Whereas if one were to vary the parameters and select values with lower calculated DMC energy, the number of required energy calculations  would scale with the number of parameters used. We should point out that in our PGD evolution examples, far more CPU time was spent calculating the energy of a single nodal surface than the entire PGD process.    
	
	PGD optimization opens up a possible way to optimize atomic positions to minimize DMC energy. If one were to combine the nuclei and electrons into one walker type, then DMC imaginary time propagation combined with the PGD method should naturally relax the atomic positions. Unfortunately we could not attempt this because the electron-nuclear cusp of the particular type of $\Psi_t$ we used depends on both the atomic positions and the variational parameters, requiring the atoms be fixed. Using a $\Psi_t$ that does not have this dependence would be an interesting avenue to explore.

	\section{Summary}
	
	We  presented a general and self-reliant method using DMC and projected   gradient
	descent that optimizes the nodal surface of a trial function to minimize the fixed-node ground state, i.e. the DMC energy.  We  derived three methods for  calculating the DMC energy gradient from  walker distribution (methods A, B, and C).
	\iffalse
	We tested these methods
	with a single trial function parameter, and method  C  produced  a  correct
	derivative to within 1\%, while methods A and B produced a
	derivative proportional to it but less by
	about half. \fi
	Methods B and C required comparable CPU time in our test, while method A was more expensive.  
	
	We  combined   the three methods  with  several   gradient  descent
	algorithms  and   a  projection operation that  maintains   the  cusp
	condition.  We benchmarked this projected  gradient descent method to trial functions with
	randomized parameters of Be, Li$_2$, and Ne. Their energies were lowered to the same level as VMC optimized parameters, without the use of a Jastrow factor.
	
	Our work  is a proof of  concept using simple systems,  but we see no reason it cannot be applied to larger and periodic  systems, or be implemented into  existing DMC code
        for various types of node determining wavefunctions.  However, before attempting these goals, our work  may be
	further improved with better and more adaptive  choices for various parameters, by simultaneously optimizing Jastrow parameters during PGD, by better suppressing  PGD fluctuations,
        and by establishing criteria for convergence.  In addition, we see no reason why the method could not be extended  to DMC on a discrete lattice or to finite temperature path integral Monte Carlo.

	\section{Acknowledgments}
	We would like to thank the Florida State University high performance computing center for the allocation of CPU time to carry out the calculations presented in this work.
	
	\appendix
	\label{section:Appendix}

	\section{Parameter fluctuations}
	\label{section:Fluctuations}	
	After  a number  of PGD  iterations, the  parameters will
	continue to fluctuate around a  local minimum due to stochastic
	error of  the energy gradient.  Here, we  investigate the  variance of
	these fluctuations. First let us perform a coordinate shift so
	that the parameters are zero at  the local minima, then let us
	rotate them so that  the Hessian $\frac{\partial^2 E}{\partial
		\theta_i \partial \theta_j}$ is  diagonal. Then, if we expand
	$E$ to second  order in $\theta_i$, one  iteration of gradient
	descent using Eq. \ref{eqn:GradientDescent} will result in new parameters $\theta_i'$ given by
	\begin{equation}
	\label{eqn:stochasticGradientDescent}
	\theta'_i = \theta_i -  a_i\big{(} K_i\theta_i + \sigma_i
	\eta\big{)} ,
	\end{equation}
	where $K_i \equiv  \frac{\partial^2 E}{\partial \theta_i^2 }$,
	$\sigma_i$ is the standard  deviation of stochastic error, and
	$\eta$  is a  random  number  with $\big{<}\eta\big{>}=0$  and
	$\big{<}\eta^2\big{>}=1$.

	Let us take the variance of both sides of
	Eq.~\ref{eqn:stochasticGradientDescent}
	\begin{equation}
	\big{<}\theta'^2_i\big{>}    =   \big{<}\theta^2_i\big{>}    +
	a^2_i\big{(}  K^2_i\big{<}\theta^2_i\big{>} +  \sigma^2_i
	\big{)} - 2a_i K_i \big{<}\theta^2_i\big{>} \; .
	\end{equation}
	After  enough PGD iterations, the variance  of fluctuations  will
	become constant, i.e.,
	\begin{equation}
	\label{eqn:constantVariance}
	\big{<}\theta'^2_i\big{>} = \big{<}\theta^2_i\big{>}.
	\end{equation}
	If this is the case, we can use Eq. \ref{eqn:stochasticGradientDescent} to evaluate the variance of latter PGD iterations, namely,
	\begin{equation}
	\label{eqn:parameterVariance}
	\big{<}\theta_i^2\big{>} = \frac{a_i  \sigma_i^2}{2 K_i -
		a_i K_i^2} \: .
	\end{equation}
	
	We  notice that the  variance blows  up if  any $a_i$  approach
	$2/K_i$, which  is the  result of  overshooting the  minimum by
	more than  double the  distance. For $a_i<<2/K_i$  we see
	that the variance has the property
	\begin{equation}
	\big{<}\theta_i^2\big{>} \propto a_i \sigma_i^2 \: .
	\end{equation}
	
	\section{Gradient descent algorithms}
	\label{section:Algorithms}

	We  list the  gradient descent  algorithms we  tested. We use
	$ g_i \equiv \frac{\partial E}{\partial\theta_i} $ as the gradient,   and
	determine $\alpha$ empirically.  We sometimes decrease $\alpha$ with iteration number to suppress parameter fluctuations around the minimum.

	%$c$,  $\beta$, $\beta_1$,  and
	%$\beta_2$ empirically. The algorithms are

	\vspace{0.15 in}

	Basic gradient descent:
	\begin{equation}
	\label{eqn:basic}
	\theta_i \rightarrow \theta_i - \alpha g_i,
	\end{equation}
	
	%	\vskip 0.15 in Gradient descent with decay:
	%	\begin{equation}
	%	\label{eqn:recipricalDecay}
	%	\begin{split}
	%	t  \rightarrow  t  +  1 \\  \theta_i  \rightarrow  \theta_i  -
	%	\frac{ \alpha g_i }{1+c^t}
	%	\end{split}
	%	\end{equation}
	
	%	\vskip 0.15 in 
	%	ADAGrad:
	%	\begin{equation}
	%	\label{eqn:ADAGrad}
	%	\begin{split}
	%	v_i \rightarrow v_i + g_i^2 \\ \theta_i \rightarrow \theta_i -
	%	\frac{\alpha g_i }{\sqrt{v_i}} \\
	%	\end{split}
	%	\end{equation}

	\vspace{0.15 in}
	
	RMSprop:
	\begin{equation}
	\label{eqn:RMSprop}
	\begin{split}
	v_i  \rightarrow  \beta  v_i    + (\beta-1) g_i^2  \\  \theta_i
	\rightarrow \theta_i - \frac{\alpha g_i }{\sqrt{v_i}} \\
	\end{split}
	\end{equation}

	\vspace{0.15 in}

	ADAM:
	\begin{equation}
	\label{eqn:ADAM}
	\begin{split}
	t  \rightarrow t  + 1  \\  m_i \rightarrow \beta_1 m_i   +
	(\beta_1-1) g_i \\ v_i \rightarrow \beta_2 v_i  +  (\beta_2-1) g_i^2
	\\  \theta_i  \rightarrow  \theta_i  -  \frac{\alpha  m_i  g_i
	}{(1-\beta_1^t) \sqrt{v_i/(1-\beta_2^t)}} \\
	\end{split}
	\end{equation}
	
	%		Max:
	%	\begin{equation}
	%	\label{eqn:ADAM}
	%	\begin{split}
	%	t  \rightarrow t  + 1  \\  m_i \rightarrow \beta m_i   +
	%	(\beta-1) g_i 
	%	\\ v_i \rightarrow \max{(v_i , |g_i|) }
	%	\\  \theta_i  \rightarrow  \theta_i  -  \frac{\alpha  m_i  g_i
	%	}{(1-\beta^t) v_i } \\
	%	\end{split}
	%	\end{equation}

	\vspace{0.15 in}

	\section{Effective $\alpha$}
	\label{section:EffectiveAlpha}

\begin{figure}[!ht]
		\centering
		\includegraphics[width=1\linewidth]{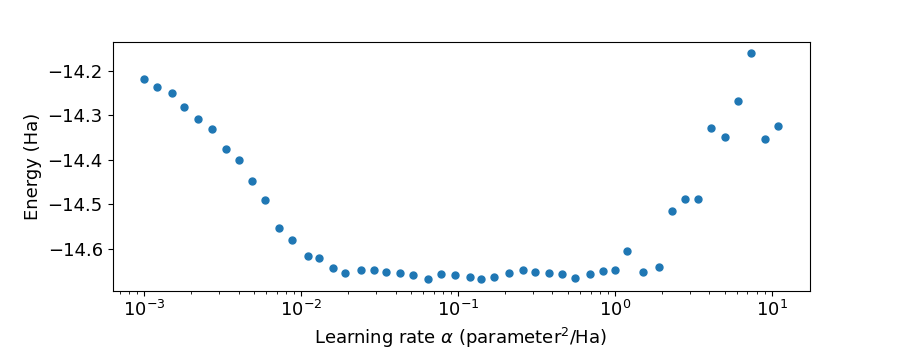} % EnergyVsAlpha.png
		\caption{An example of how the energy, calculated after a 100 PGD iterations, depends on different learning rates $\alpha$. This data was used for the 6th row of Table~\ref{table:table}. }
		% This calculation used a Be trial-function with randomized parameters and the PGD used RMSprop with $\beta=0.99$ and method C with $\tau_c = 0.25$ Ha$^{-1}$. 
		\label{fig:plot7}
	\end{figure}

	In Table. \ref{table:table} we show the effective ranges of the learning rate $\alpha$, which is a hyper parameter proportional to a shift in parameters, as shown in Appendix~\ref{section:Algorithms}.
	We define this effective range as the values that bring the energy of a randomized $\Psi_S$ down to the optimum level, and show an example in Fig. \ref{fig:plot7}.  We empirically determine this range for 100 PGD iterations for different combinations of gradient descent algorithms, atomic systems, and methods A, B or C. We display the end-points of the effective range along with the
	logarithmic distance between them.

	Compared to basic gradient descent, the adaptive algorithms RMSprop and ADAM have more consistent endpoints for methods A, B and C and for different atomic systems, making the choice of $\alpha$ easier. We attribute the more consistent endpoints to the division of each gradient component by a trailing root square mean of its prior values. This suppresses the large variations of gradient magnitudes for different atomic systems. It also suppresses large variations of $\xi(R,\tau_c)$ of Eq. \ref{eqn:methodC} used for method C, which has a factor of $\exp(\Delta E \; \tau_c)$, with $\Delta E $ the difference between estimated and correct energy.

	\begin{table}[!h]
		\centering
		\begin{tabular}{ |c|c|c|c|c|c| }
			\hline
			Algorithm 
			&  \begin{tabular}{l}   Gradient  \\ method \\ \end{tabular} 
			& $\Psi_S$ %&  Min $\alpha_{eff}$ &  Max $\alpha_{eff}$
			&  \begin{tabular}{l}   Minimum  \\ effective $\alpha$ \\ \end{tabular}   
			&  \begin{tabular}{l}  Maximum  \\ effective $\alpha$ \\ \end{tabular}   
			&  $\log\Big{(}\frac{  \alpha_{ \mbox{\tiny max} }}{\alpha_{ \mbox{\tiny min} }}\Big{)}$ \\ \hline
			\multirow{3}{*}{ \begin{tabular}{l}   Basic  \\ gradient \\ descent\\ \end{tabular}   } 
			& B & Be                          & $1.2\times10^{-1}$   & $1.3\times10^{0}$\:\:\:& $1.0$  \\ 
			& C & Be                          & $2.4\times10^{-2}$   & $6.4\times10^{-1}$     & $1.4$  \\ 
			& A & Li$_2$                      & $4.8\times10^{-3}$   & $2.0\times10^{-1}$     & $1.6$  \\  
			& B & Ne                          & $4.5\times10^{-2}$   & $1.8\times10^{0}$\:\:\:& $1.6$  \\ 
			\hline
			\multirow{3}{*}{RMSprop} & B & Be & $5.5\times10^{-2}$   & $1.0\times10^{0}$\:\:\:& $1.3$  \\  
			& C & Be                          & $2.0\times10^{-2}$   & $1.0\times10^{-1}$     & $1.7$  \\ 
			& A & Li$_2$                      & $1.8\times10^{-2}$   & $2.1\times10^{3}$\:\:  & $5.1$  \\ 
			& B & Ne                          & $5.6\times10^{-2}$   & $1.5\times10^{0}$\:\:\:& $1.4$  \\ 
			\hline
			\multirow{3}{*}{ADAM} & B & Be    & $3.8\times10^{-2}$   & $1.0\times10^{0}$\:\:\:& $1.4$  \\ 
			& C & Be                          & $1.1\times10^{-2}$   & $1.0\times10^{0}$\:\:\:& $2.0$  \\ 
			& A & Li$_2$                      & $1.6\times10^{-1}$   & $7.0\times10^{2}$\:\:  & $3.6$  \\  
			& B & Ne                          & $2.4\times10^{-1}$   & $9.2\times10^{0}$\:\:\:& $1.6$  \\ 
			\hline
			
		\end{tabular}
		\caption{The range of the effective learning rate $\alpha$ for 100 iterations for three gradient descent algorithms: basic gradient descent (common $a_i$), RMSprop, and ADAM (see Fig. \ref{fig:plot7} for an example of the tenth row). RMSprop used $\beta=0.99$. ADAM used $\beta_1=0.9$, $\beta_2=0.99$. Method C used $\tau_c = 0.3$ Ha$^{-1}$}
		\label{table:table}
		
	\end{table}

	\FloatBarrier

%aapmrev4-2.bst 2019-01-14 (MD) hand-edited version of aapmrev4-1.bst

\end{document}